%% file: arxiv.tex
\documentclass[runningheads]{llncs}

\usepackage[T1]{fontenc}
\usepackage{graphicx}
\usepackage{subcaption}

\usepackage{booktabs}
\usepackage{multirow}
\usepackage{arydshln}
\usepackage[normalem]{ulem}
\useunder{\uline}{\ul}{}
\usepackage{siunitx}
\usepackage{enumitem}

\usepackage{algorithmicx}
\usepackage{algorithm}
\usepackage[noend]{algpseudocode}
\usepackage{listings}

\usepackage{hyperref}
\usepackage{color}

\usepackage{comment}

\begin{document}

\title{From Formulas to Figures: How Visual Elements Impact User Interactions in Educational Videos}
\titlerunning{From Formulas to Figures}

\author{Wolfgang Gritz\inst{1}\orcidID{0000-0003-1668-3304} \and
Hewi Salih\inst{2} \and
Anett Hoppe\inst{1,3,4}\orcidID{0000-0002-1452-9509} \and
Ralph Ewerth\inst{1,3,4}\orcidID{0000-0003-0918-6297}}

\institute{TIB -- Leibniz Information Centre for Science and Technology, %, Welfengarten 1B, 30167,
Hannover, Germany \email{wolfgang.gritz@tib.eu} \\
\and
Leibniz University Hannover, %30167, 
Hannover, Germany \and
L3S Research Center, Leibniz University Hannover, %Appelstraße 9A, 30167, 
Hannover, Germany \and
University of Marburg and hessian.AI  -- Hessian Center for Artificial Intelligence, %Hans-Meerwein-Straße 6, 35032, 
Marburg, Germany
}

\authorrunning{W. Gritz et al.}

\maketitle              % typeset the header of the contribution

\begin{abstract}
Educational videos have become increasingly relevant in today's learning environments.
While prior research in laboratory studies has provided valuable insights, analyzing real-world interaction data can enhance our understanding of authentic user behavior.
Previous studies have investigated technical aspects, such as the influence of cuts on pausing behavior, but the impact of visual complexity remains understudied.
In this paper, we address this gap and propose a novel approach centered on visual complexity, defined as the number of visually distinguishable and meaningful elements in a video frame, such as mathematical equations, chemical formulas, or graphical representations.
Our study introduces a fine-grained taxonomy of visual objects in educational videos, expanding on previous classifications.
Applying this taxonomy to \num{25} videos from physics and chemistry, we examine the relationship between visual complexity and user behavior, including pauses, in-video navigation, and session dropouts.
The results indicate that increased visual complexity, especially of textual elements, correlates with more frequent pauses, rewinds, and dropouts.
The results offer a deeper understanding of how video design affects user behavior in real-world scenarios.
Our work has implications for optimizing educational videos, particularly in STEM fields.
We make our code publicly available\footnote{\url{https://github.com/TIBHannover/from_formulas_to_figures}}.

\end{abstract}

%
% ---- Introduction ----
%
\section{Introduction}\label{sec:intro}
Educational videos have become a central component of modern education, allowing learners to control their learning pace and interact with the material.
The rise of online education platforms and the digital transformation of traditional teaching formats has increased their importance~\cite{navarrete2021review,navarrete2023closer,Sablic2021Review}.
Videos are particularly effective in enhancing comprehension and retention since they combine visual and auditory elements, as highlighted in the Cognitive Theory of Multimedia Learning (CTML)~\cite{mayer2014multimedialearning}.
However, while the CTML provides a robust framework for understanding the benefits of multimedia learning, its key principles still require further testing and validation in real-world settings~\cite{Mayer2024Future}.

While there is extensive research on the effects of pausing, less attention has been paid to the motivation behind user interactions~\cite{Biard2018Effects,Hasler2007control,Liu2021complexity-determined,Schwan2004Dynamic,Spanjers2012Explaining}.
Furthermore, most of the research refers to controlled laboratory conditions and only a few to realistic interaction data~\cite{Anders2024Mindwandering,Merkt2022Pushing}.
In this regard, the role of visual complexity (i.e., the number of meaningful elements) remains understudied.
Understanding the relationship between visual complexity and user behavior is essential, as excessive visual complexity could increase extraneous cognitive load, making it difficult for learners to process information.

This paper aims to shed light on the relationship between visual complexity and its influence on user interactions (i.e., pausing, rewinding, forwarding, and dropping out).
To explore this relationship, we first developed a taxonomy for visual objects in the context of science and technology education (STEM).
Visually complex content, such as formulas, diagrams, and graphs, is ubiquitous in these subjects and might pose particular challenges.
We annotated \num{25} videos from a channel with physics, chemistry, and math videos from an open-source video platform\footnote{\url{https://av.tib.eu/}} and gathered corresponding interaction data.
Our statistical evaluation relies on a permutation-based approach.
We adopted the Dynamic Time Warping (DTW) algorithm for the test statistic to capture the interplay of visual complexity and possibly delayed user interactions.
We aim to answer the following research question:

\begin{quote}
\emph{RQ: How does visual complexity in STEM learning videos influence user interactions such as pausing, seeking, and dropping out?}
\end{quote}

The remainder of the paper is structured as follows:
Section~\ref{sec:rw} comprises research related to our work.
In Section~\ref{sec:dataset}, we present the creation of our dataset, including the selection of videos and the development of the taxonomy for visual objects, as well as the annotation process.
Section~\ref{sec:data_processing} details the methodology, explaining our data processing approach and the modified DTW algorithm.
In Section~\ref{sec:results}, we discuss the results of our analysis, highlighting key findings on the relationship between visual complexity and user interactions. Finally, Section~\ref{sec:conclusions} concludes the paper summarizing our findings.

%
% ---- Related Work ----
%
\section{Related Work}\label{sec:rw}
In this section, we first summarize recent advances in video learning research regarding user interactions (Section~\ref{sec:rw:interactions}).
Subsequently, we highlight research on visual complexity in educational videos (Section~\ref{sec:rw:viscom}).

\subsection{User Interactions}\label{sec:rw:interactions}
One key principle of CTML is the segmenting effect, which suggests that learning can be improved when multimedia content is broken down into smaller, manageable segments rather than presented as a continuous stream.
This might prevent cognitive overload by allowing learners to process information in chunks, making integrating new knowledge into their existing mental models easier.
In their meta-analysis, Rey et al.~\cite{Rey2019meta} concluded that learners could benefit from dividing videos into segments.
Specifically, they found that learners with higher prior knowledge derive greater advantage from segmentation as they can use the additional processing time to integrate new information with their existing knowledge base more effectively.
This leads to better retention and transfer performance.
Spanjers et al.~\cite{Spanjers2012Explaining} found two different explanations for this phenomenon: (1)~the slight pauses between the segments help to reduce cognitive load~\cite{Lusk2009Multimedia,Spanjers2011Reversal}, and (2)~the segmentation itself cues the learning of the underlying structure~\cite{Yu2024Cues}.
By segmenting the content, learners might be less likely to become overwhelmed, and they can focus more effectively on understanding each part.

Pauses can be distinguished into learner-determined pauses and system-determined pauses.
Learner-determined pauses allow individuals to control the pace and choose when to stop the video according to their needs.
However, recent studies found that learners barely use the pause function~\cite{Biard2018Effects,Hasler2007control}.
On the other hand, system-determined pauses are typically predefined according to the structure of the learning material~\cite{Merkt2018Pauses}; however, they can also be implemented at other points, such as around particularly complex sections~\cite{Liu2021complexity-determined}, to guide learners through challenging content.
Liu et al.~\cite{Liu2021complexity-determined} placed system-determined pauses before and after sections with high element interactivity within educational videos.
They found that these pauses increased learning outcomes and efficiency compared to a control group viewing continuous videos (without pauses).
However, they observed no significant difference in learning outcomes between the group using system-determined pauses and the group using learner-determined pauses.
While most previous studies~\cite{Biard2018Effects,Hasler2007control,Liu2021complexity-determined,Schwan2004Dynamic,Spanjers2012Explaining} have been conducted in controlled lab settings, research using authentic usage data has been limited.
Some studies analyzed authentic viewing patterns to identify user types~\cite{Akcapinar2024Decoding,Guo2014production}, but few explore what makes learners interact in real-world settings~\cite{Anders2024Mindwandering,Merkt2022Pushing}.
Merkt et al.~\cite{Merkt2022Pushing} analyzed the influence of human-annotated meaningful structure, difficulties in comprehension, and structural breakpoints derived from an automatic shot boundary detection algorithm.
They found a connection between pausing behavior and meaningful structure but not with the structural breakpoints.
Anders et al.~\cite{Anders2024Mindwandering} also found a relationship between pausing behavior and meaningful structure, and additionally with mind-wandering.

\subsection{Visual Complexity in Educational Videos}\label{sec:rw:viscom}
Visual complexity, in general, can be described as the level of detail within an image~\cite{donderi2006viscom,forsythe2009viscom}.
Perceptual processes often simplify this complexity, as numerous details (e.g., pixels) may be grouped into familiar patterns based on prior knowledge.
Individual elements such as labels or mathematical terms could be perceived as fundamental when applied to educational videos with few visual changes (e.g., slide-based videos).
Meier et al.~\cite{Meier2023socialcues} studied the influence of social cues under different conditions of visual complexity and found increased learning outcomes in the low-complexity setting.
They controlled visual complexity by adding textual labels to the videos.
Similarly, van der Zee et al.~\cite{vanderZee2017effects} conceptualize the quantity of text in a video as a key dimension of visual-textual information complexity.
However, in real scenarios, various visual objects can appear in educational videos.
In this study, \textit{visual complexity} refers to the number of visually distinguishable objects or elements in a video frame.
These elements can be textual (e.g., mathematical terms) and graphical (e.g., images or diagrams).

%
% ---- Approach ----
%
\section{Development of the Dataset}\label{sec:dataset}
This section comprises our data collection process, describing the video platform (Section~\ref{sec:platform}), our taxonomy creation and data annotation process (Section~\ref{sec:taxonomy}), and a summary of the gathered study data (Section~\ref{sec:study_data}).

\subsection{Video Platform and Data Collection}\label{sec:platform}
We use the open and non-commercial platform TIB AV-Portal\footnotemark[2] as our video source.
The TIB AV-Portal's collection of over \num{40000} scientific videos spans various fields, focusing strongly on documentaries and educational materials from natural sciences and technology.
User interaction with videos, such as starting, pausing, resuming, seeking, and finishing, is tracked within the platform if the user consents\footnote{https://av.tib.eu/privacy}.
The data for this study was anonymized.

\begin{figure}[t]
    \centering
    \includegraphics[width=0.69\linewidth]{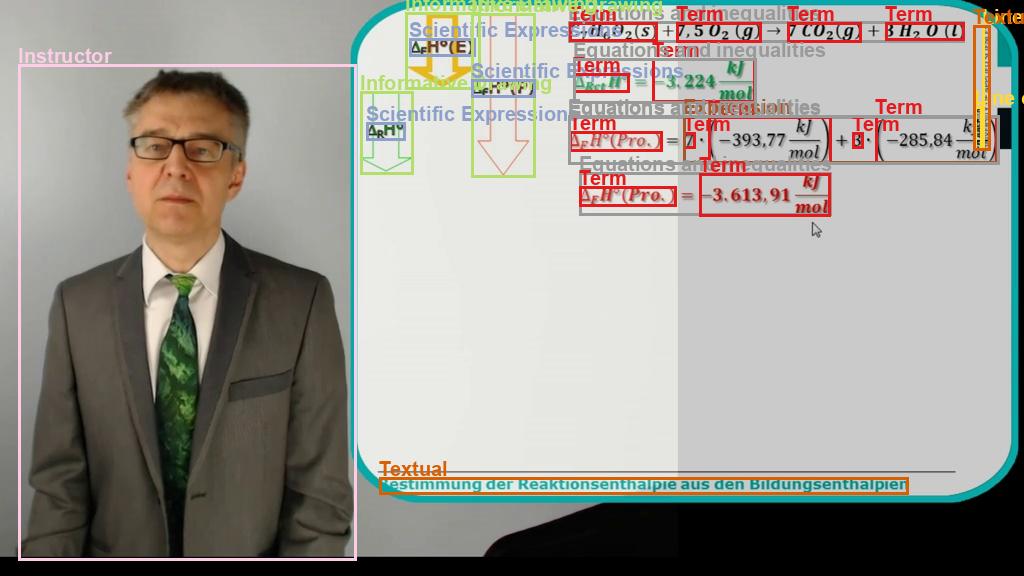}
    \caption{Annotated example frame (see 4:46 at \url{https://av.tib.eu/media/15698})}
    \label{fig:example_frame}
\end{figure}

This study aimed to investigate the influence of visual complexity on user interaction.
Therefore, we selected videos that contain many visual objects of different categories.
Furthermore, a high volume of collected user interactions is important for the significance of the results.
Therefore, we selected videos from the SciFox channel, one of the platform's most popular channels that garners numerous views and user interactions.
The videos on this channel primarily deal with chemistry topics but have large overlaps with physics and mathematics. 
Figure~\ref{fig:example_frame} shows an example frame. 
The composition of the SciFox videos usually includes the speaker and overlays of visual objects such as data visualizations and mathematical terms.

\subsection{Taxonomy for Visual Objects and Annotation Process}\label{sec:taxonomy}
\begin{figure}[t]
    \centering
    \includegraphics[width=\textwidth]{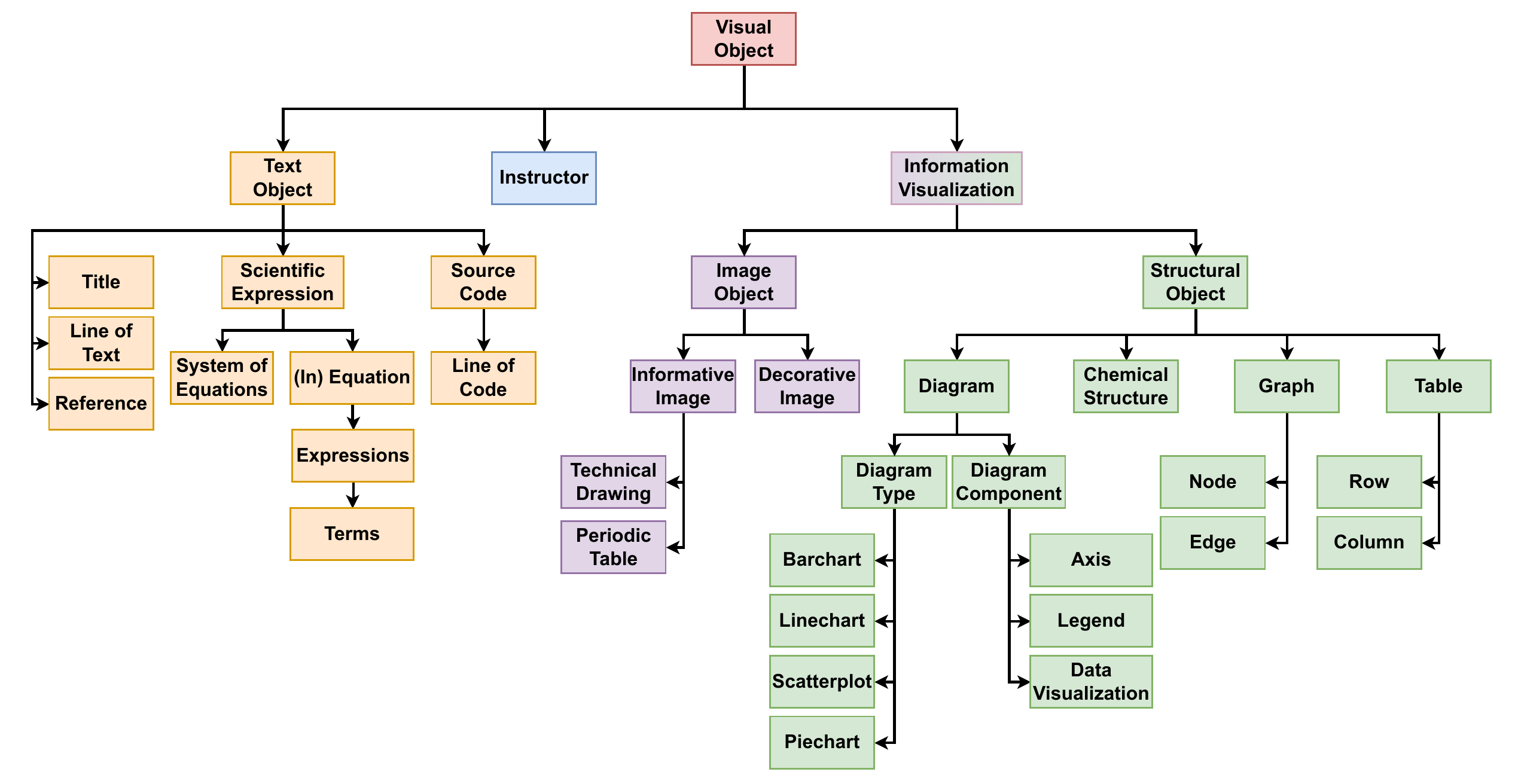}
    \caption{Shows the visualization of the developed taxonomy for visual objects in chemistry, physics, and mathematics videos.}
    \label{fig:taxonomy}
\end{figure}

\subsubsection{Inspiration and Background:}
Our taxonomy for visual objects in educational videos is inspired by existing datasets for object detection in slide-based learning videos~\cite{SPaSe2019,Wise2019,FitVid2022,Seng2022Enriching}.
While these datasets provide valuable insights, they have different foci and lack the detail required for our specific needs.
For instance, we aimed to capture nuanced information, such as distinguishing between equations with two versus five mathematical terms.
Furthermore, we used the taxonomy for key terms in math education by Castro \& Gomez~\cite{castro2021taxonomy}.

\subsubsection{Development Process:}
Our objective was to create a taxonomy that is both comprehensive and compact.
A computer science student ($A_1$) and a second person with a computer science degree ($A_2$) employed an iterative, exploratory approach to developing this taxonomy.
Reviewer guidelines were established and continuously adapted throughout the process.
When an object could not be assigned, we either revised the guidelines or modified the taxonomy.

\subsubsection{Annotation Methodology:}
For each visual object, (1) a rectangular bounding box was annotated, and (2) tracking IDs were assigned to monitor changes between frames.
These IDs help differentiate between newly appearing objects and those in previous frames.
The annotation was primarily performed by $A_1$ using the open source tool \texttt{cvat}\footnote{\url{https://www.cvat.ai/}}.
Five diverse frames were selected and annotated by $A_2$ to ensure the annotation quality.
We measured the agreement between annotators using the Intersection over Union (IoU) metric, which reflects the overlap in position, size, and assigned class of the bounding boxes.
The taxonomy and guidelines were modified until we achieved a sufficient IoU value of \num{0.88}.
After reaching this level of agreement, $A_1$ completed the annotation for all videos.

\subsection{Extraction of Video Frames}
Since the timestamps at which a visual change occurs are relevant, we aimed to identify these and extract the corresponding frames.
We used \texttt{FFmpeg}\footnote{\url{https://www.ffmpeg.org/}} with the built-in scene filter.
We carefully calibrated individual thresholds for each video to ensure no visual changes were missed.
Redundant frames (i.e., without anything appearing or disappearing) were removed manually. 

\subsection{Study Data}\label{sec:study_data}
We selected \num{25} videos with lengths of 5:01 to 11:26 minutes.
For the \num{25} videos, \num{4185} unique objects (\num{15368} in total) were annotated, comprising \num{3481} textual objects and \num{704} information visualization objects.
We collected anonymous log files spanning from August 23, 2018, to April 5, 2023, encompassing between \num{74} to \num{1363} individual user sessions with an average of \num{327}, and \num{8180} in total consisting of at least one "play" event.
In these \num{8180} sessions, the users paused the videos \num{40012} times, rewinded \num{15308} times, and skipped \num{34350} times.
Since the log data might contain user sessions with interaction data from multiple videos during a browser session, we cleaned possible redundancy between logs.

\section{Processing of the Data}\label{sec:data_processing}
This section summarizes our methodology to investigate the relationship between visual complexity and user interactions.
Therefore, we first describe the conceptualization of visual complexity (Section~\ref{sec:conception_viscom}).
Next, we explain the data preprocessing steps (Section~\ref{sec:preprocessing}).
Finally, we introduce our statistical approach to measure the significance and strength of effects (Section~\ref{sec:statistical_analysis}).

\subsection{Conceptualization of Visual Complexity}\label{sec:conception_viscom}
As motivated in Section~\ref{sec:rw:viscom}, we formulate visual complexity as the change in the number of various objects across consecutive frames.
Specifically, we use the tracking IDs of bounding boxes appearing in the current frame but not in the previous frame.
This approach allows us to quantify the frequency and intensity of visual changes.
We recompute the results with binarized values to validate whether it is not the magnitude of the change (i.e., the level of visual complexity) but rather the mere occurrence of a visual change that has an effect.

Since we hypothesize that user behavior might differ for text-based or image-based elements, we categorized class labels as \textbf{textual}, \textbf{information visualization} according to our taxonomy (see Section~\ref{sec:taxonomy}) as follows:
\begin{itemize}
    \item \textbf{Textual Elements}: Text Object, Title, Line of Text, Scientific Expressions, Systems of Equations, Terms
    \item \textbf{Information Visualization Elements}: Information Visualization, Image Object, Informative Image, Technical Drawing, Structural Object, Barchart, Linechart, Scatterplot, Diagram Component, Axis, Data Visualization, Chemical Structure, Table, Row, Column
\end{itemize}
This also ensures that visual elements are not counted multiple times in the analysis (e.g., three mathematical terms forming an equation); they are included only as three individual terms.
On the other hand, the varying number of components in charts or tables is covered.
This methodology allows for a granular and differentiated analysis, avoiding overlaps and redundancies.

\subsection{Preprocessing of Time Series}\label{sec:preprocessing}
\begin{figure}[t]
    \centering
    \includegraphics[width=1.0\linewidth]{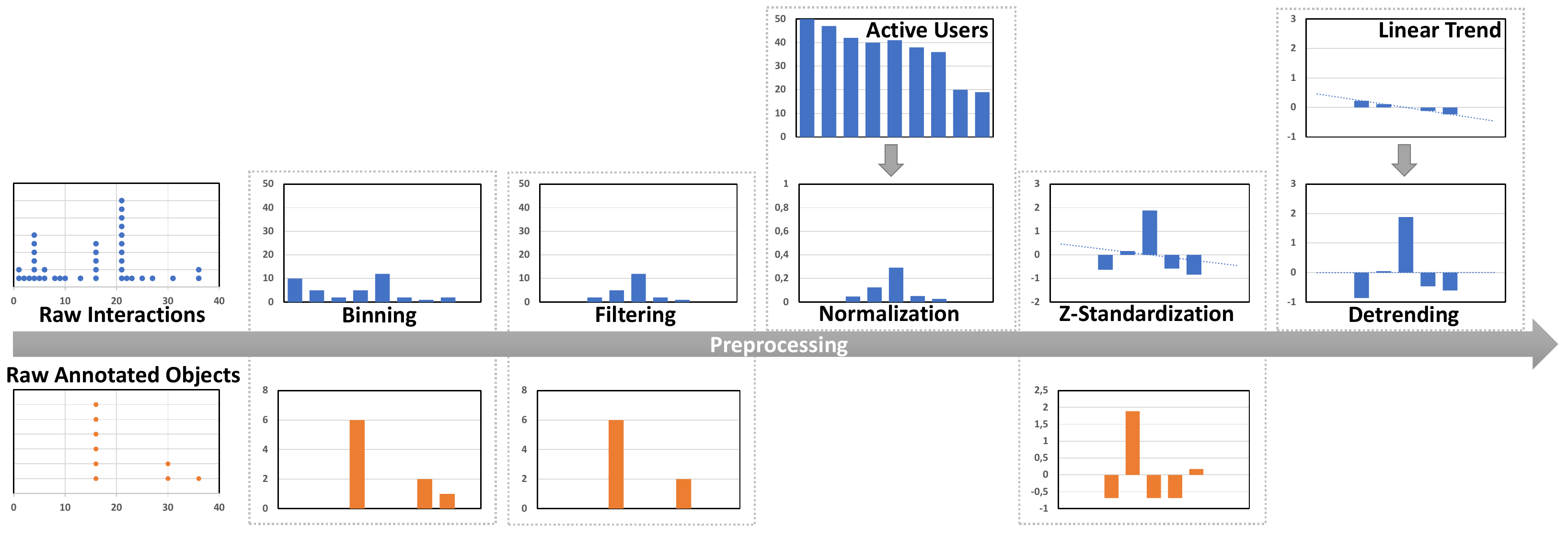}
    \caption{Overview of our data preprocessing (Section~\ref{sec:preprocessing}).}
    \label{fig:overview_preprocessing}
\end{figure}
This section comprises the data preprocessing as visualized in Figure~\ref{fig:overview_preprocessing}.

\subsubsection{Binning Procedure}\label{sec:binning}
We first convert the continuous event data (user interactions) and the visual complexity data into discrete time series by creating histograms with fixed bin sizes of five seconds, following the literature~\cite{Anders2024Mindwandering,faber2018driven}.
The selection seems reasonable, as in our case, a smaller window chosen by Merkt et al.~\cite{Merkt2022Pushing} could be too short for the users' reaction time.
The binning process results in two parallel time series for each video: one representing user interactions (i.e., pauses, rewinds, skips, and dropouts) and the other representing visual complexity.
Note: For each bin, the unique sessions with interactions were counted, not the interactions themselves, to avoid bias caused by users pausing several times in the same \qty{5}{\second} time frame.

\subsubsection{Data Filtering}\label{sec:filtering}
When working with authentic usage data, the input data may contain noise and unwanted effects.
Therefore, we applied two filtering steps:
\begin{itemize}
    \item \textbf{Active session selection:} We filtered out sessions where users left the page before starting the video, i.e., we filtered all sessions without a "play" event).
    \item \textbf{Edge case handling:}
    Frequent pauses and rewinds occur at the very start of the video, likely due to users adjusting to the content (e.g., the video starts automatically).
    Additionally, we observe a sharp decline in user interactions near the end of the video when the closing remarks are shown.
    Therefore, we excluded the first two and last two bins from the analysis.
\end{itemize}

\subsubsection{Data Normalization}
Given the varying number of users across different videos and bins, we applied preprocessing steps to the time series data as follows:
\begin{itemize}
    \item \textbf{Normalization by active sessions:}
    To account for varying user counts over time, the number of sessions with user interactions was normalized by the total number of active sessions within each bin.
    \item \textbf{Z-normalization:}
    The visual complexity and user interaction time series were z-normalized for each video.
    After applying this preprocessing step, both time series now have a mean of \num{0} and a standard deviation of \num{1}, enabling a proper comparison between them.
    \item \textbf{Detrending:}
Since the frequency of user interactions (e.g., pauses or rewinds) may increase or decrease as the video progresses, we applied a linear detrending procedure using \texttt{scikit-learn}~\cite{scikit-learn} to the user interaction time series to remove potential biases introduced by long-term trends.
\end{itemize}

\subsection{Statistical Analysis}\label{sec:statistical_analysis}
Previous research methods, such as localized analysis of isolated events by Anders et al.~\cite{Anders2024Mindwandering} and direct time series comparisons by Merkt et al.~\cite{Merkt2022Pushing}, were deemed unsuitable for our study.
Unlike previous work, our data features frequent and consecutive visual stimuli.
This means that an analysis of the local environment like Anders et al.~\cite{Anders2024Mindwandering} was unsuitable since there are relatively few instances where isolated visual changes occur.
Based on exploratory visual analysis, we observed that user responses (e.g., pausing) often occur up to \num{10} seconds after a visual stimulus, but with variable delays rather than a consistent fixed time lag.
A direct time series comparison, as used by Merkt et al.~\cite{Merkt2022Pushing}, or a cross-correlation analysis would not accommodate these variable temporal delays between visual stimuli and user reactions.
As a result, we performed a permutation test with a custom test statistic that allows for flexible temporal delays in user reactions.
The custom test statistic based on the dynamic time warping algorithm and determination of significance and effect sizes are described below.

\subsubsection{Modified Dynamic Time Warping (DTW) Algorithm}\label{sec:dtw}
\input{algo/dtw_algorithmicx}
We utilized a modified version of the DTW algorithm to align the time series of visual complexity with the time series of user interactions.
DTW is commonly used in signal processing to measure the similarity between two sequences that may vary in speed or timing.
It minimizes the cumulative cost of aligning the two sequences, defined as the squared difference between each point in the visual complexity series and the corresponding point in the user interaction series.
However, standard DTW allows bidirectional alignment.
In our case, this would mean assigning user interactions to future visual stimuli, which is unsuitable.
Therefore, we modified the algorithm with two key constraints:

\begin{itemize}
    \item \textbf{Causality constraint:} To ensure that user interactions are only mapped to preceding visual complexity changes, we set all costs below the main diagonal of the distance matrix to infinity.
    This prevents future user interactions from being mapped to past visual complexity changes.
    \item \textbf{Reaction window constraint:} Based on exploratory analysis, we expect user reactions to occur within a maximum of \num{10} seconds after a stimulus.
    To enforce this, we set all costs for points more than two bins ahead of the visual stimulus to infinity.
    This limits the warping window and ensures that mappings are localized to a reasonable temporal range.
\end{itemize}

The modified DTW algorithm is detailed in Algorithm~\ref{alg:dtw}.
This algorithm produces a temporal constrained mapping of visual complexity changes to user interactions and a total cost for the mapping for each video.

\subsubsection{Significance and Effect Size}
We implemented a permutation testing approach to assess the statistical significance of the alignment between visual complexity and user interaction.
For every video $v$, we repeated following procedure:
\begin{enumerate}
    \item First, we calculate the test statistic ($\text{cost}_{v}$) for the original visual complexity time series ($\text{VisCom}_{v}$) with the user interaction time series ($\text{UserInt}_{v}$), based on the modified DTW algorithm.
    
    \item Next, we generate a null distribution ($\text{Perm}_{v}$) by permuting $\text{VisCom}_{v}$ randomly for $N_{\text{Perm}_{v}}=\num{5000}$ times and calculating the test statistic for each permutation with $\text{UserInt}_{v}$.
    
    \item Subsequently, we calculate the proportion of permutations in $\text{Perm}_{v}$ that resulted in extremer (i.e., lower resp. higher) test statistics compared to the original sequence.
    Subsequently, the significance value $p_{v}$ is defined as:
    \begin{equation}
        p_{v} = \frac{\#\{ |\text{Perm}_{v} - \mu_{\text{Perm}_{v}}| \geq |cost_{v} - \mu_{\text{Perm}_{v}}| \}+1}{N_{\text{Perm}_{v}}+1},
    \end{equation}
    where $\mu_{\text{Perm}_{v}}$ denotes the mean.
    \num{1} is added to ensure that $p_v\neq \num{0}$.
    
    \item Finally, we assess the strength of relationships using a measure similar to Cohen's $d$, a standard effect size metric.
    We define the \textit{permutation effect size (PES)} as the standardized difference between $cost_{v}$ and the mean permuted cost $\mu_{\text{Perm}_{v}}$, using its standard deviation $\sigma_{\text{Perm}_{v}}$:
    \begin{equation}
        \text{PES}_{v} = - \frac{cost_{v} - \mu_{\text{Perm}_{v}}}{\sigma_{\text{Perm}_{v}}}.
    \end{equation}
    The negative sign ensures a positive $\text{PES}_{v}$ indicates a positive.
\end{enumerate}
This results in \num{25} p-values and \num{25} values for PES, that we aggregate as follows:
\begin{enumerate}[resume]
    \item To aggregate the \num{25} p-values to $p$, we use Fisher's method~\cite{Fisher1992}.
    
    \item The \num{25} $\text{PES}_{v}$ values are averaged to obtain PES, reflecting the overall effect size and capturing positive and negative relationships.
    
    \item Our definition of the p-values allows positive or negative relationships, which is not considered in Fisher's method.
    Therefore, we compute the \qty{95}{\percent} confidence intervals (CI) using bootstrapping across the \num{25} $\text{PES}_{v}$ values as an additional indicator for statistical significance.
    In particular, we consider an effect significant if both conditions are fulfilled: (1)~$p<\num{0.05}$ and (2)~$0\notin \text{CI}$.
\end{enumerate}
Finally, we report PES and CI alongside the significance values to provide a more comprehensive understanding of the strength of the relationships.

\section{Results}\label{sec:results}
\input{tables/significance_table}

Our results (see Table~\ref{table:significance}) show that visual complexity has a significant influence on pausing behavior (p=\num{0.009}, PES=\num{0.64}).
In particular, users paused more frequently in sections with a high number of textual changes (p=\num{0.0}, PES=\num{0.89}).
This pattern aligns with Cognitive Load Theory~\cite{paas2003cognitiveloadtheory}, which posits that complex information increases intrinsic cognitive load, prompting users to take additional time to process text.
Interestingly, no significant relationship between a high number of visualizations and user pauses could be found.
This might suggest that while textual complexity increases cognitive demands, visualizations may be processed more fluidly, reducing the need for pauses.
This supports the CTLM~\cite{mayer2014multimedialearning}, which proposes that well-designed visuals engage the visual processing channel and reduce extraneous load, facilitating continuous engagement.
For textual changes, the binary condition still showed a significant effect (p=\num{0.047}, PES=\num{0.57}).
This highlights that the degree of visual complexity, particularly the textual changes, is more influential than simply the occurrence of a change.

Furthermore, we also observed a significant relationship between visual complexity and user dropout from a video for textual elements (p=\num{0.017}, PES=\num{0.47}).
The lack of a significant effect in the binary setting leads to the conclusion that videos are not abandoned for arbitrary changes but rather when there are many textual elements.
This could be due to cognitive overload and loss of engagement.

We observed a similar pattern with rewinds.
Users were more likely to rewind after encountering high visual complexity (p=\num{0.017}, PES=\num{0.64}), especially for textual elements (p=\num{0.001}, PES=\num{0.79}), while we observed no effect in the binary condition.
However, no significant effect was observed for visualizations regarding rewinds, indicating that users are more likely to re-watch sections with textual complexity rather than sections dominated by visual content.
Similar to the dropout observation, there was no significant effect in the binary settings, supporting the hypothesis that users have difficulty at textually complex points in the video.
It is also worth noting that while visual complexity might trigger rewinding, it seems independent of the target time of rewinding.
Finally, we did not observe a significant effect between visual complexity and skipping behavior at the source or target times.

The results suggest that textual elements induce more pausing, rewinding, and dropout behavior than visual content.
The increased pausing and rewinding around text-heavy sections may indicate higher cognitive load and users taking additional time to process the content or abandoning the video.
Based on the observed effects of visual complexity on user interactions, three key recommendations emerge for enhancing educational content delivery.
First, a visual complexity assessment tool for content creators would empower them to analyze and visualize the complexity of their materials before publication.
Such a tool's comprehensive textual and visual complexity metrics can help creators balance information delivery with user engagement to improve the educational experience.
Second, integrating visual cues~\cite{Spanjers2012Explaining,Yu2024Cues} within video players highlighting complex segments could enable learners to take effective pauses.
This proactive approach could encourage more profound interaction with challenging content and promote active learning.
Lastly, enriching videos at complex points with interactive elements, such as questions or supplementary information, could enhance viewer comprehension and retention~\cite{jing2016interpolated}.

%
% ---- Conclusions ----
%
\section{Conclusions}\label{sec:conclusions}
This paper analyzes the relationship between visual complexity in educational videos and user interactions, such as pauses and rewinds.
Therefore, we first developed a taxonomy for visual objects in STEM educational videos.
Next, we annotated \num{25} videos and gathered corresponding real-world interaction data from an open video platform.
Finally, we performed a permutation-based approach using an adopted Dynamic Time Warping algorithm for statistical analysis.

Our results showed increased pausing, rewinding, and dropout behaviors following increased visual complexity based on textual elements.
This observation supports the idea that text demands more cognitive resources in learning videos.
The increased dropouts suggest that users were generally less willing to continue through visually complex sections and, therefore, consider abandoning the video.
In contrast, visual elements did not increase these behaviors significantly, indicating that users might not find visual content as cognitively demanding as textual information.
However, our results are based on \num{25} videos, and the findings' generalizability requires further research.

Our results highlight the importance of balancing textual and visual information in educational videos to optimize user engagement.
For content creators, tools that assess and visualize visual complexity could help design more effective educational materials.
Additionally, highlighting potentially complex segments in video timelines or adding small interactive elements to these sections could further enhance user understanding and engagement.

\begin{credits}
\subsubsection{\ackname}
Part of this work was financially supported by the Leibniz Association, Germany (Leibniz Competition 2023, funding line "Collaborative Excellence", project VideoSRS [K441/2022]).

\subsubsection{\discintname}
The authors have no competing interests to declare that are relevant to the content of this article.
\end{credits}

%
% ---- Bibliography ----
%
\bibliographystyle{splncs04}
\bibliography{bibliography}

\end{document}

%% file: algo/dtw_algorithmicx.tex
\begin{algorithm}[tb]
\caption{Modified Dynamic Time Warping (DTW) Algorithm}
\label{alg:dtw}
\begin{algorithmic}[1] 
    \Require Two sequences \( s_1 \) and \( s_2 \)
    \Require \texttt{window}: Size of the warping window
    
    \State Initialize distance matrix \( D \) with shape \( (len(s_2), len(s_1)) \)
    
    \For{each \( i \) in \( 1 \) to \( len(s_2) \)}
        \For{each \( j \) in \( 1 \) to \( len(s_1) \)}
            \State \( D[i, j] = (s_1[j] - s_2[i])^2 \)
        \EndFor
    \EndFor

    \State Set distance matrix outside of window to \( \infty \)

    \State Initialize cost matrix \( C \) with shape \( (len(s_2), len(s_1)) \)
    \State \( C[0, 0] = D[0, 0] \)
    
    \For{each \( i \) in [\( 1 \), ..., \( len(s_2) \)]}
        \State \( C[i, 0] = D[i, 0] + C[i-1, 0] \)
    \EndFor
    
    \For{each \( j \) in [\( 1 \), ..., \( len(s_1) \)]}
        \State \( C[0, j] = D[0, j] + C[0, j-1] \)
    \EndFor
    
    \For{each \( i \) in [\( 1 \), ..., \( len(s_2) \)]}
        \For{each \( j \) in [\( 1 \), ..., \( len(s_1) \)]}
            \State \( C[i, j] = D[i, j] + \min(C[i-1, j], C[i, j-1], C[i-1, j-1]) \)
        \EndFor
    \EndFor
    
    \State Calculate total cost as \( \text{total\_cost} = \frac{C[len(s_2)-1, len(s_1)-1]}{len(path)} \)

    \State \textbf{return} \( \text{total\_cost} \), optimal path
\end{algorithmic}
\end{algorithm}

%% file: tables/significance_table.tex
\begin{table}[t]
    \setlength{\tabcolsep}{3.9pt}
    \centering
    %\fontsize{9}{10}\selectfont
    \caption{Comparison of p-values (p) and permutation effect sizes (PES) and corresponding confidence intervals (CI) for binary visual changes and visual complexity (VisCom) across user actions and visual (V) and textual (T) modalities (mod). Significant results are bold ($p<0.05$) and underlined ($0\notin CI$).}
    \label{table:significance}
    \begin{tabular}{llccccccc}
    \toprule
     &  & \multicolumn{1}{l}{} & \multicolumn{3}{c}{\textbf{Binary}} & \multicolumn{3}{c}{\textbf{VisCom}} \\
    \textbf{action} & \textbf{time} & \textbf{mod} & \textbf{p} & \textbf{PES} & \textbf{CI} & \textbf{p} & \textbf{PES} & \textbf{CI} \\
    \midrule
    \multirow{3}{*}{\textbf{pause}} & \multirow{3}{*}{\textbf{}} & \textbf{T} & {\ul \textbf{0.047}} & {\ul \textbf{0.57}} & {[}0.14, 1.00{]} & {\ul \textbf{0.0}} & {\ul \textbf{0.89}} & {[}0.39, 1.28{]} \\
     &  & \textbf{V} & \textbf{0.003} & -0.19 & {[}-0.70, 0.33{]} & 0.410 & -0.48 & {[}-0.84, -0.13{]} \\
     &  & \textbf{T+V} & 0.266 & 0.48 & {[}0.07, 0.80{]} & {\ul \textbf{0.009}} & {\ul \textbf{0.64}} & {[}0.16, 1.05{]} \\
     \midrule
    \multirow{3}{*}{\textbf{dropout}} & \multirow{3}{*}{\textbf{}} & \textbf{T} & \textbf{0.045} & 0.14 & {[}-0.34, 0.61{]} & {\ul \textbf{0.017}} & {\ul \textbf{0.47}} & {[}0.02, 0.94{]} \\
     &  & \textbf{V} & \textbf{0.009} & -0.16 & {[}-0.67, 0.32{]} & 0.940 & -0.23 & {[}-0.49, 0.05{]} \\
     &  & \textbf{T+V} & 0.099 & 0.18 & {[}-0.28, 0.60{]} & 0.112 & 0.37 & {[}-0.07, 0.79{]} \\
     \midrule
    \multirow{6}{*}{\textbf{rewind}} & \multirow{3}{*}{\textbf{from}} & \textbf{T} & 0.386 & 0.63 & {[}0.32, 0.94{]} & {\ul \textbf{0.001}} & {\ul \textbf{0.79}} & {[}0.30, 1.24{]} \\
     &  & \textbf{V} & 0.166 & 0.14 & {[}-0.28, 0.55{]} & 0.634 & -0.18 & {[}-0.51, 0.16{]} \\
     &  & \textbf{T+V} & 0.48 & 0.61 & {[}0.27, 0.85{]} & {\ul \textbf{0.017}} & {\ul \textbf{0.64}} & {[}0.20, 1.07{]} \\
     & \multirow{3}{*}{\textbf{to}} & \textbf{T} & 0.319 & 0.24 & {[}-0.17, 0.59{]} & 0.327 & 0.44 & {[}0.08, 0.85{]} \\
     &  & \textbf{V} & 0.664 & -0.06 & {[}-0.44, 0.26{]} & 0.360 & 0.0 & {[}-0.40, 0.40{]} \\
     &  & \textbf{T+V} & 0.282 & 0.33 & {[}-0.09, 0.67{]} & 0.260 & 0.44 & {[}0.09, 0.88{]} \\
     \midrule
    \multirow{6}{*}{\textbf{skip}} & \multirow{3}{*}{\textbf{from}} & \textbf{T} & 0.146 & -0.50 & {[}-0.95, -0.10{]} & 0.085 & -0.26 & {[}-0.73, 0.19{]} \\
     &  & \textbf{V} & 0.326 & -0.50 & {[}-0.87, -0.13{]} & 0.875 & -0.03 & {[}-0.32, 0.31{]} \\
     &  & \textbf{T+V} & 0.478 & -0.35 & {[}-0.72, 0.03{]} & 0.154 & -0.17 & {[}-0.61, 0.26{]} \\
     & \multirow{3}{*}{\textbf{to}} & \textbf{T} & 0.116 & -0.18 & {[}-0.64, 0.24{]} & 0.729 & 0.19 & {[}-0.15, 0.54{]} \\
     &  & \textbf{V} & \textbf{0.014} & -0.39 & {[}-0.91, 0.06{]} & 0.247 & 0.06 & {[}-0.32, 0.49{]} \\
     &  & \textbf{T+V} & 0.212 & -0.11 & {[}-0.59, 0.28{]} & 0.699 & 0.22 & {[}-0.12, 0.58{]} \\
     \bottomrule
    \end{tabular}
\end{table}